\documentclass[aps,twocolumn,superscriptaddress,floatfix,longbibliography]{revtex4-1}
\usepackage{tikz}
\usepackage{lipsum}
\usepackage{graphicx}
\usepackage[export]{adjustbox}
\usepackage{amsmath,amssymb,wasysym}
\usepackage{braket}
\usepackage{mathtools}
\usepackage{mathrsfs}
\usepackage{verbatim}
\usepackage{makecell}
\usepackage{cases}
\usepackage{tikz}
\usepackage[compat=1.1.0]{tikz-feynman}
\usepackage{ulem}

\usepackage{slashed}
\usepackage{hyperref}
\usepackage{xcite}
\usepackage{xr}

\usepackage{xcolor}
\usepackage{pgfplots}
\pgfplotsset{compat=1.18}
\makeatletter
\newcommand{\im}{\mathbf{i}}

\renewcommand{\Im}{\mathrm{Im}}

\newcommand{\BigO}{\mathcal{O}}

\def\Q{\mathbf{Q}}

\def\k{\mathbf{k}}
\def\q{\mathbf{q}}
\def\x{\mathbf{x}}

\begin{document}
\title{Linear Magnetoresistance from Glassy Orders}
\author{Jaewon Kim}
\affiliation{Department of Physics, University of California, Berkeley, CA 94720, USA}  
\author{Ehud Altman}
\affiliation{Department of Physics, University of California, Berkeley, CA 94720, USA}
\author{Shubhayu Chatterjee}
\affiliation{Department of Physics, Carnegie Mellon University, Pittsburgh, PA 15213, USA}

\begin{abstract}
Several strongly correlated metals display B-linear magnetoresistance (LMR) with a universal slope, in sharp contrast to the $B^2$ scaling predicted by Fermi liquid theory. 
We provide a unifying explanation of the origin of LMR by focusing on a common feature in their phase diagrams --- proximity to symmetry-breaking orders.
Specifically, we demonstrate via two microscopic models that LMR with a universal slope arises ubiquitously near ordered phases, provided the order parameter either (i) has a finite wave-vector, or (ii) has nodes on the Fermi surface.
We elucidate the distinct physical mechanisms at play in these two scenarios, and derive upper and lower bounds on the field range for which LMR is observed. 
Finally, we discuss possible extensions of our picture to strange metal physics at higher temperatures, and argue that our theory provides an understanding of recent experimental results on thin film cuprates and moir\'e materials.
\end{abstract}
\maketitle

\begin{figure}[!htbp]
    \centering
    \includegraphics[width = 0.9\columnwidth]{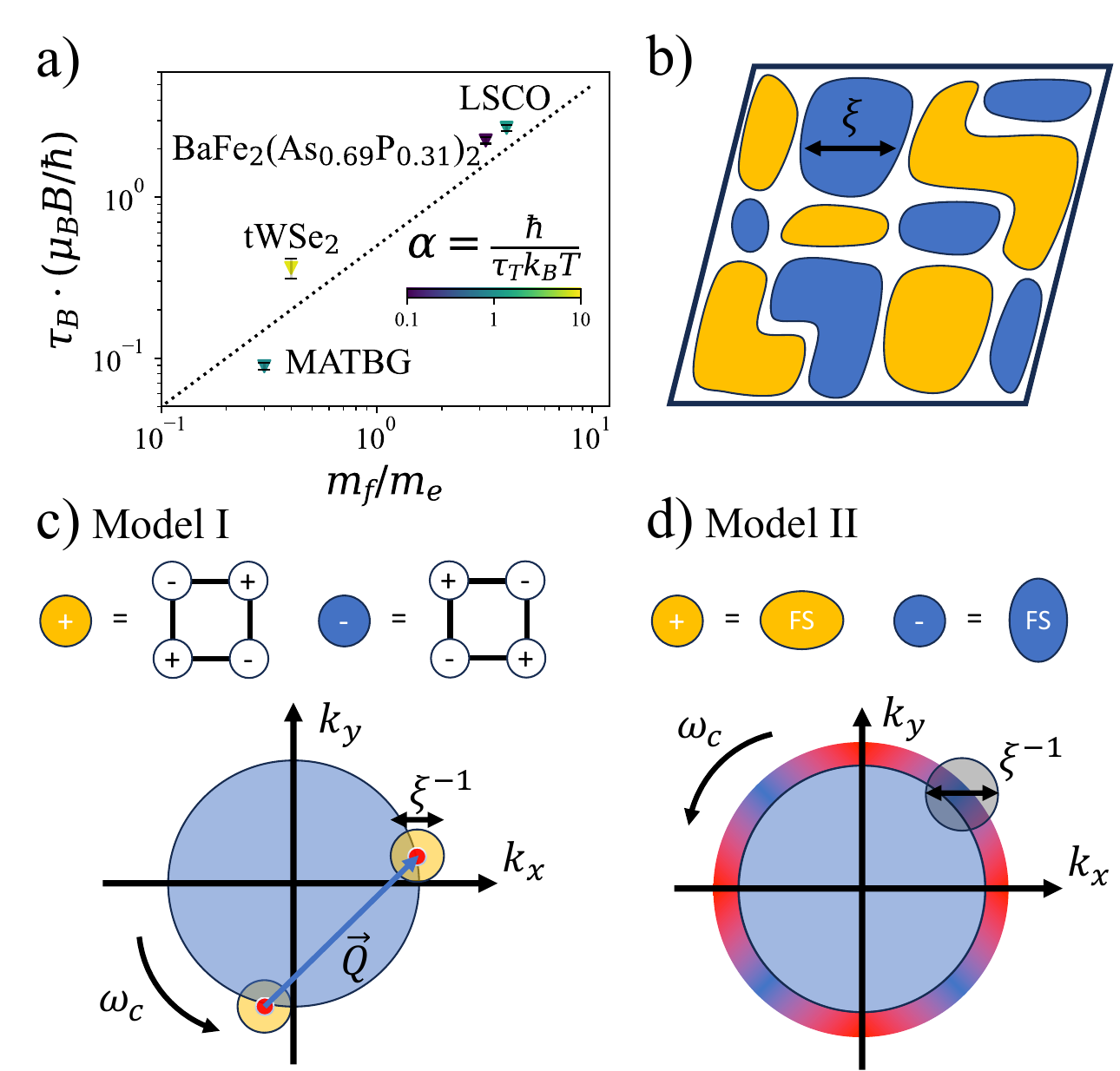}
    \caption[(a) Quasi-particle relaxation time $\tau_{B}$ inferred from a Drude analysis \cite{LSCO,Hayes,Ghiotto,Jaoui} scaled by $\mu_B B/\hbar$, versus the effective mass $m_f/m_e$ measured via quantum oscillations \cite{LSCOmass1,Pnictidemass,Ghiotto,Jaoui} for different materials exhibiting LMR. 
    Note that $\tau_B (\mu_B B/\hbar) \approx m_f/m_e$ for all these materials, indicating a universal relaxation rate set by $\tau_B^{-1} \approx \tilde{\mu}_B B/\hbar$.
    (b) A schematic depiction of our physical setting: quenched order parameter domains of typical size $\xi$ that couple to electrons near the Fermi surface. Crucially for LMR, we require that the order parameter has either a finite wave-vector $\Q$ (Model I) or nodes on the FS (Model II).
    (c) In model I, $B$ rotates quasi-particles into hot-spots of size $\xi^{-1}$ (yellow circles) where they strongly backscatter, resulting in an overall relaxation rate of $\BigO(\omega_c)$. (d) 
    In model II, $B$ removes the momentum relaxation bottleneck, caused by weak scattering at the cold-spots (blue regions) on the FS, by rotating quasiparticles across them, which leads to a faster relaxation rate of $\BigO(\omega_c)$.]
    {(a) Quasi-particle relaxation time $\tau_{B}$ inferred from a Drude analysis \cite{LSCO,Hayes,Ghiotto,Jaoui} scaled by $\mu_B B/\hbar$, versus the effective mass $m_f/m_e$ measured via quantum oscillations \cite{LSCOmass1,Pnictidemass,Ghiotto,Jaoui} for different materials exhibiting LMR \footnotemark. 
    Note that $\tau_B (\mu_B B/\hbar) \approx m_f/m_e$ for all these materials, indicating a universal relaxation rate set by $\tau_B^{-1} \approx \tilde{\mu}_B B/\hbar$.
    (b) A schematic depiction of our physical setting: quenched order parameter domains of typical size $\xi$ that couple to electrons near the Fermi surface. Crucially for LMR, we require that the order parameter has either a finite wave-vector $\Q$ (Model I) or nodes on the FS (Model II).
    (c) In model I, $B$ rotates quasi-particles into hot-spots of size $\xi^{-1}$ (yellow circles) where they strongly backscatter, resulting in an overall relaxation rate of $\BigO(\omega_c)$. (d) 
    In model II, $B$ removes the momentum relaxation bottleneck, caused by weak scattering at the cold-spots (blue regions) on the FS, by rotating quasiparticles across them, which leads to a faster relaxation rate of $\BigO(\omega_c)$.
    }
    \label{fig:headimage}
\end{figure}

\textit{Introduction.}--
Metals are ubiquitous in nature, and it is commonly thought that most of their transport properties can be explained through semiclassical Boltzmann theory \cite{mah00,Ashcroft76}.
However, a wide array of strongly correlated quantum materials displays, in their metallic phase, puzzling transport properties that violate the basic tenets of the standard theory \cite{Phillips,Greene,Chowdhury2}.
One persistent puzzle is the observation of linear magnetoresistance (LMR), i.e., $\Delta \rho(B) = \rho(B) - \rho(0) \propto B$, in a variety of (quasi-) two-dimensional correlated electronic materials, such as cuprates \cite{2009Sci...323..603C,LSCO}, pnictides \cite{Hayes}, 
iron chalcogenides \cite{FeSe}, and most recently in moir\'e systems \cite{Ghiotto,Jaoui,Unmesh}.
Such behavior is in stark contrast to the prediction of the semiclassical theory - $\Delta \rho(B) \propto B^2$ \cite{lifshitz1956theory,ziman1958galvanomagnetic,pippard1989magnetoresistance}. 
Curiously, many of these materials exhibit LMR in the proximity of symmetry-breaking orders in their phase diagrams.
Further, the quasi-particle relaxation rate $\tau_B^{-1}$ inferred via a Drude analysis of transport measurements satisfies $\tau_B^{-1} \approx \tilde\mu_B B/\hbar$ with $\tilde{\mu}_B = e \hbar/m_f$ \footnote{$\tilde\mu_B$ denotes the renormalized Bohr magneton, $\frac{e \hbar}{m_f}$, where $m_f$ denotes the quasi-particle mass}, independent of the specific material (Fig.~\ref{fig:headimage}a) \cite{LSCO,Hayes,Ghiotto,Jaoui}.
The pervasive occurrence of unusual magnetic field scaling and universal relaxation rate raises an intriguing puzzle. 

Several theoretical proposals have attempted to explain these observations \cite{Parish,Aavishkar,Gil,2022NatPh..18.1420A,Hussey,Feng,Koshelev2,Maksimovic,Rosch,Koshelev1,Zou}.
In particular, Ref.~\cite{Hussey} proposed a possible unifying perspective: LMR can arise from impeded cyclotron motion of quasiparticles in the vicinity of the Fermi surface (FS). 
Specifically, the authors used the relaxation time approximation to argue that if certain regions of the Fermi surface have very short relaxation times, they effectively obstruct cyclotron motion leading to LMR.
While this physical picture is quite appealing, it immediately leads to several further questions. 
Which microscopic models lead to such anisotropic relaxation and concurrently  exhibit the universal relaxation rate?
What sets the upper and lower bounds on the magnetic field required for LMR?
What happens when forward scattering dominates, rendering the relaxation time approximation ineffective \cite{Ashcroft76}?

\footnotetext{The data is also represented as a table in the SI.}

Here, we demonstrate that B-linear magnetoresistance with universal relaxation arises ubiquitously in the proximity of ordered electronic phases.
To this end, we consider concrete microscopic models of electrons coupled to glassy symmetry breaking orders with frozen domains, where the order parameter either has (i) a finite wave-vector $\Q$ (e.g., charge-density wave order) or (ii) nodes on the Fermi surface (e.g., nematic order). 
In scenario (i), the presence of glassy density-wave order leads to the formation of hot-spots on the Fermi surface connected by wave-vector $\Q$, where quasiparticles strongly back-scatter. 
Once the magnetic field is turned on, the quasiparticles can rotate into the hot-spots at a rate proportional to the cyclotron frequency $\omega_c = eB/m_f$, leading to LMR.
By contrast, in scenario (ii), a nodal order parameter leads to cold-spots on the Fermi surface where quasiparticle scattering is strongly suppressed, leading to a bottleneck in relaxation of the non-equilibrium momentum distribution. 
The magnetic field now rotates electrons out of the cold-spots at a rate $\omega_c$, leading to removal of this bottleneck and consequently LMR.
Notably, in both cases the quasi-particle relaxation rate inferred from electrical conductance is set by the cyclotron frequency $\omega_c$ when LMR is observed, and follows $\tau_B^{-1} \approx \tilde\mu_B B/\hbar$, in accordance with experiments.
Our microscopic model also allows us to establish concrete upper and lower bounds on magnetic fields for which LMR is expected. 
Finally, assuming Planckian dissipation, i.e., $\tau_T^{-1} \approx k_B T/\hbar$ at zero field  \cite{Bruin804}, we estimate the crossover temperature scale from quadratic to linear magnetoresistance at finite temperatures. 
We argue that such crossover can happen at low magnetic fields $\tilde \mu_B B \ll k_B T$ --- an observation \cite{LSCO} that cannot be explained by phenomenological resistor network models of LMR \cite{Parish,Aavishkar}.

\textit{LMR from glassy density waves.-}
We consider spinless electrons occupying a two dimensional Fermi sea, coupled to uncorrelated potential disorder and to glassy (static) density wave order with wave-vector $\Q$ correlated over a lengthscale $\xi$.
This is captured by the following Hamiltonian (model I).
\begin{subequations}
\label{eq:dpipiL}
\begin{align}
    &\mathcal{H} = H_f + H_{d} + H_{dw} \,, \textrm{ where } \nonumber \\
    &H_f = \int_{\k} \left(\epsilon_\k - \mu \right) c_\k^\dagger  c_\k
    \label{eq:Hf} \\
    &H_{d} = \sum_\x g_\x c^\dagger_\x c_\x
    \label{eq:Hdd} \\
    &H_{dw} = \sum_\x J e^{\im \Q \cdot \x} n_\x c_\x^\dagger c_\x + \mbox{H.c}.
    \label{eq:Hdis1}
\end{align}
\end{subequations}
For simplicity, in \eqref{eq:Hf} we consider a quadratic dispersion $\epsilon_\k = \frac{\k^2}{2m_f}$ \footnote{Henceforth, we set $\hbar = 1$.}.
The potential disorder in \eqref{eq:Hdd} has zero mean, $\overline{g_\x} = 0$ and is uncorrelated, $\overline{g_\x g_{\x^\prime}} = g^2 \delta_{\x \x^\prime}$; it leads to an isotropic relaxation rate $\tau_{\mu}^{-1} = m_f g^2$ \cite{mah00}.
The final ingredient of our model, in \eqref{eq:Hdis1} is glassy density-wave order, which is responsible for anisotropic scattering of electrons and plays a crucial role in LMR. 
Specifically, we consider a scenario with $\overline{n_\x} = 0$ (no long-range density-wave order) and $\overline{n_\x n_{\x^\prime}} = e^{-|\x - \x^\prime|^2/4\xi^2}$, with a correlation length $\xi$ much larger than the microscopic lattice spacing $a$, such that $k_F \xi \gg 1$ \footnote{While we consider Gaussian correlated glassy order, the precise form is correlations is not as important as the fact that it is correlated over lengthscales of $\xi$. We show this in the SI by explicitly recovering our main results when the order parameter correlations has a different distribution.}. 

To understand the effect of the glassy density wave order on the low energy quasiparticles on the Fermi surface, it is instructive to consider $H_{dw}$ in momentum space.
\begin{align}
H_{dw} = \sum_{\k,\q} J n_\q c^\dagger_{\k+\Q+\q} c_\k + \mbox{H.c.}
\label{eq:HdwMomentum}
\end{align}
In \eqref{eq:HdwMomentum}, $n_\q$ is the Fourier transform of $n_\x$ and satisfies $\overline{n_\q} = 0, \ \overline{n_\q n_{\q^\prime}} = \pi \xi^2 e^{-\xi^2 q^2}\delta_{\q+\q^\prime}$.
We may now use Fermi's golden rule to calculate $\Gamma_{\k_i \rightarrow \k_f}$, the average rate at which a quasiparticle at momentum $\k_i$ scatters to momentum $\k_f$ by the glassy order.
\begin{equation}
\begin{split}
    \Gamma_{\k_i \rightarrow \k_f} = \frac{2\pi^2 J^2 \xi^2}{v_F} e^{-\xi^2 q^2} \,, \textrm{ where } \q = \k_f - \k_i \pm \Q
    \label{eq:GammaCDW}
\end{split}
\end{equation}
Next, we evaluate $\tau_{\k_i}^{-1}$, the scattering rate of a quasiparticle with initial momentum $\k_i$. $\tau_{\k_i}^{-1} = \int_{k_f} \Gamma_{\k_i \rightarrow \k_f}$; it is largest at the points of the Fermi surface which are connected by wavevector $\Q$ -- often called hot-spots.
In the weak-coupling limit $J \lesssim v_F/\xi$, the scattering rate at a hot-spot is given by
\begin{equation}
    \frac{1}{\tau_{h}} \simeq \frac{2\pi^2 J^2 \xi}{v_F} \lesssim \frac{v_F}{\xi}
    \label{eq:SigmaCDW}
\end{equation}
Note that the back-scattering rate is upper-bounded by $v_F/\xi$, which is the time scale for a quasiparticle to cross an ordered domain and scatter from a domain wall. 
For an initial momentum $\k_i$ separated by $\q$ from the nearest hot-spot, the scattering rate is suppressed exponentially by $e^{-\xi^2 q^2}$.
Ergo, the backscattering processes from the density wave disorder are localized to a few hotspots of size $\BigO(1/\xi)$ on the Fermi surface (Fig.~\ref{fig:headimage}b).

Having derived the scattering rates at the hot-spots, we are ready to understand the physical mechanism that leads to LMR in the presence of glassy density waves. 
Since the hotspots only occupy a small fraction of order $\BigO(1/k_F \xi)$ of the Fermi surface, most quasi-particles do not feel the effect of $H_{dw}$ when the magnetic field $B$ is absent.
Accordingly, the relaxation rate is set by the isotropic decay rate from the potential disorder $\tau_\mu^{-1}$.
This leads to a zero-field resistivity given by the Drude formula $\rho_0 \simeq \left( \frac{m_f}{ne^2}\right) \tau_\mu^{-1}$.
However, once a magnetic field is turned on, the quasi-particles start rotating around the Fermi surface at the cyclotron frequency $\omega_{c} = \frac{eB}{m_f}$.
When quasi-particles enter a hotspot, they strongly back-scatter to another hotspot via a large momentum transfer $\Q$, leading to impeded cyclotron motion \cite{Hussey} and relaxation of current.
As a result, the average quasi-particle lifetime becomes proportional to the time required to rotate into a hotspot, which scales as the time-period of the cyclotron orbit.
Consequently, we find a B-linear magnetoresistance,
\begin{equation}
    \rho(B) \simeq \left( \frac{m_f}{ne^2} \right) (\tilde\mu_B B)\,,
    \label{eq:rhoBuniv}
\end{equation}
with a decay rate that follows the universal relation, $\tau_B^{-1} \simeq \tilde\mu_B B$.

To confirm our semi-classical predictions, we numerically solve the Boltzmann equation in the DC limit
(see SI for a derivation).
\begin{equation}
\begin{split}
    &\frac{e v_F \hat{\k} \cdot \mathbf{E}}{2\pi} + \omega_{c} \nabla_{\theta_k} \delta N_{\hat{\k}} = - \frac1{\tau_\mu} \delta N_{\hat{\k}} \\
    & + \int_{\hat{\k'}} \frac{\pi J^2}{\epsilon_F} (k_F \xi)^2 e^{-\xi^2 |k_F(\hat{\k}- \hat{\k'}) \pm \Q|^2} \Big(\delta N_{\hat{\k'}} - \delta N_{\hat{\k}} \Big) 
    \label{eq:BEmod_1}
\end{split}
\end{equation}
Here, we have restricted to angular coordinates $\hat{\k}$ on the Fermi surface, such that $\delta N_{\hat{\k}}$ denotes the deviation from equilibrium quasiparticle density in the $\hat{\k}$ direction, and fixed $\Q = \frac{\pi}{a}(1,1)$ for concreteness.
The left hand side of \eqref{eq:BEmod_1} constitute the electromagnetic force experienced by the quasi-particles; the right hand side comprises the collision integral and represents the collision processes undergone by the quasi-particles: the first term describes scattering from potential disorder, and the second term accounts for scattering from glassy density waves with the rate given in \eqref{eq:GammaCDW}.

\begin{figure}[ht]
    \centering
    \includegraphics[width = 1 \columnwidth]{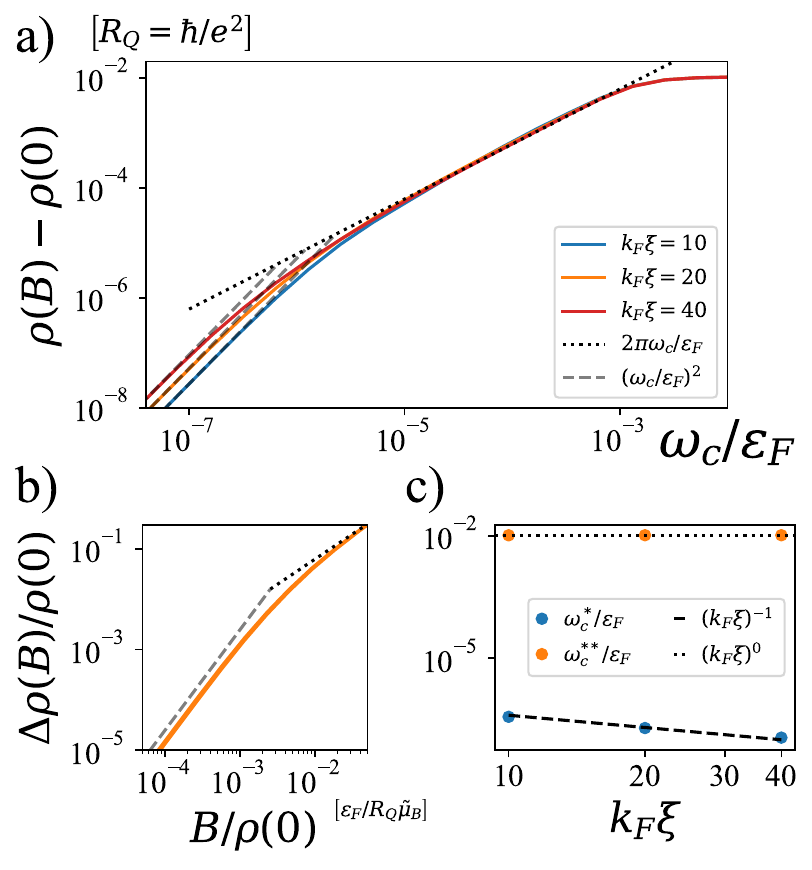}
    \caption{(a) $\Delta \rho(B) = \rho(B)-\rho(0)$ as a function of $\omega_c$ for different density-wave correlation lengths $\xi$, obtained by numerically solving \eqref{eq:BEmod_1} for $\tau_\mu^{-1} = 2 \times 10^{-5} \epsilon_F$ and $J = 0.025 \epsilon_F.$
    In the linear regime $\rho(B) - \rho(0) \simeq 2\pi \omega_c/\epsilon_F = \frac{m_f}{ne^2} \tilde\mu_B B$.
    There is a crossover from $B^2$ to $B$ linear magnetoresistance at $\omega_c^*$, and then from $B$ linear to constant at $\omega_c^{**}$.
    Here, $\omega_c^*$ and $\omega_c^{**}$ are extracted as the points where the $B^2$ (dashed line) or constant asymptote meet the $B$ linear asymptote (dotted line).
    (b) Scaling collapse that occurs when both axes are rescaled by the zero-field resistivity $\rho_0 \sim 1/{\epsilon_F\tau_\mu}$ for various values of potential scattering lifetime $\tau_\mu^{-1} \in \{10^{-5}\epsilon_F, 1.6 \times 10^{-4} \epsilon_F\}$, indicating Kohler's rule is satisfied.
    (c) Numerical evidence for power-law scaling of the lower and upper cross-over fields $\omega_c^{*}$ and $\omega_c^{**}$ with $k_F \xi$.
    }
    \label{fig:model1}
\end{figure}

In Fig.\ref{fig:model1} we plot the magnetoresistance $\Delta \rho (B)$ for various correlation lengths $\xi$.
Several features of $\Delta \rho(B)$ are immediately apparent. 
First, at intermediate fields we note the appearance of LMR, with the slope given by \eqref{eq:rhoBuniv}.
Second, panel b) shows a scaling collapse of the magnetoresistance rescaled with the zero-field resistance i.e., $\Delta \rho(B)/\rho(0)$ versus $B/\rho(0)$, remiscinent of Kohler's rule \cite{kohler1938magnetischen}.
Finally, we observe two distinct crossover behaviors: (i) from Fermi-liquid like quadratic ($B^2$) to B-linear magnetoresistance at $\omega_c^* = 1/(\tau_\mu k_F\xi)$, and (ii) a saturation of magnetoresistance to a constant value at $\omega_c^{**} = 1/(\tau_h k_F\xi)$ (Fig.~\ref{fig:model1}c).
Both of these crossovers can be explained using simple semi-classical arguments, as we elaborate below.

To understand the low-field crossover to LMR, consider a quasiparticle on the Fermi surface: 
On average, it rotates through an angle $\Delta \theta \approx \omega_{c}\tau_\mu$ before it decays due to short-range potential disorder.
Each hot-spot on the other hand occupies roughly an $\BigO(1/k_F\xi)$ angular extent on the Fermi surface.
Hence, if $\omega_{c}\tau_\mu \lesssim 1/k_F\xi$, any quasiparticle at the fringes of the hotspots will decay through scattering from the potential disorder before it reaches the heart of the hotspot, rendering current relaxation due to hot-spot scattering ineffective. 
Conversely, for $\omega_{c}\tau_\mu \gtrsim 1/k_F\xi$, the quasiparticles at the hotspot fringes reach the center and decay through scattering off glassy density waves, thus setting the crossover cyclotron frequency $\omega_c^* \propto 1/(\tau_\mu k_F \xi)$, which is also the lower bound for LMR.
We note that the lower crossover cyclotron frequency $\omega_c^*$ goes to zero as the angular extent of the hot-spot decreases ($1/(k_F \xi) \to 0$), consistent with the onset of LMR at vanishingly small magnetic fields in presence of delta-function impedance for cyclotron motion \cite{Hussey}.
However, the non-universal $\BigO(1)$ prefactor that governs $\omega_c^*$ does depend on the geometry of the hotspot scattering, such as the shape of the Fermi surface and the precise location and angular distances between the hotspots.

To understand the high-field saturation, note that that the time spent by a quasiparticle in the hot-spot region is given by $1/(\omega_c k_F \xi)$.
If the scattering time at the hot-spot $\tau_h$ is smaller than $1/(\omega_c k_F \xi)$, then the quasiparticle is able to scatter from density waves before it rotates through the hot-spot, and we get LMR.
On the other hand, if $\tau_h$ is much larger than $1/(\omega_c k_F \xi)$, the quasiparticle rotates through the hotspot without scattering, and LMR is lost.
This sets the upper bound $\omega_c^{**} = 1/(\tau_h k_F \xi)$ for LMR, beyond which $\rho(B)$ saturates.
In the weak coupling limit $J \lesssim v_F/\xi$, we may use our Fermi's golden rule result from \eqref{eq:SigmaCDW}, $1/\tau_h \sim (J^2/\epsilon_F)(k_F \xi)$ to estimate $\omega_c^{**} \sim J^2/\epsilon_F$.
Note that in the strong coupling limit $J \gtrsim v_F/\xi$ but $J \ll \epsilon_F$, we expect $\tau_h$ to saturate to $\xi/v_F$.
Accordingly, $\omega_c^{**}$ would scale as $\epsilon_F/(k_F\xi)^2$ in this limit. 

The coupling strength $J$ also plays an important role in determining the nature of quantum oscillations associated with the Shubnikov de Haas effect \cite{Ashcroft76} at magnetic fields beyond the upper crossover value.
Specifically, the Fermi surface size measured through the quantum oscillations differs: for $J \lesssim v_F/\xi$, we expect to see a single large Fermi surface; for $J \gtrsim v_F/\xi$ on the other hand, we expect several smaller Fermi pockets.
To see why, let us focus on a single domain, where the CDW order is essentially long-range and reconstructs the Fermi surface into smaller Fermi pockets.
These pockets are separated in momentum space by $\Delta k \sim J/v_F$ since $J$ is effectively the strength of the hybridization.
Crucially, since the size of a domain is $\BigO(\xi)$, the momenta are smeared by $\xi^{-1}$.
Therefore, in the weak coupling limit where $J \lesssim v_F/\xi$, the fuzziness of the momentum eliminates the pockets -- in turn, we expect to see a single, large Fermi surface.
Conversely in the strong coupling limit where $J \gtrsim v_F/\xi$, the smearing of the momentum is smaller than the momentum separation between the pockets;
in turn, we expect to probe small folded Fermi pockets through the Shubnikov-de Haas effect.

\textit{LMR from glassy nematic order.-}
We now discuss a different mechanism, where LMR arises because the magnetic field releases a bottelneck for momentum relaxation.  
Specifically, we consider electrons coupled to a nodal order parameter - glassy Ising nematic order on a square lattice (model II):
\begin{equation}
    H_{nem} = \sum_\x J n_\x \big\{c_{\x+\hat{x}}^\dagger c_\x - c_{\x+\hat{y}}^\dagger c_\x + \mbox{H.c.} \big\} \,.
    \label{eq:Hdis_2}
\end{equation}
Here $n_\x$ is quenched (Ising) nematic order with zero mean (no long-range nematic order) and Gaussian correlations over a length-scale $\xi$, as before. 
For simplicity, we first consider the limit of vanishing potential disorder, such that all electronic scattering arises from coupling to glassy nematicity.

It is convenient to write the scattering on the glassy order in momentum space
\begin{equation}
\begin{split}
    H_{nem}
    &\simeq \sum_{\k,\q} J \cos 2\theta_{\k+\frac{\q}{2}} n_\q c^\dagger_{\k+\q} c_\k \; \; \text{near the FS}.
    \label{eq:Hq_nem}
\end{split}
\end{equation}

Now we can find the scattering rate for a quasi-particle at momentum $\k_i$ using Fermi's golden rule
\begin{equation}
    \frac1{\tau_{\k_i}} \simeq \frac{J^2}{2\epsilon_F} (k_F \xi)^2 \left(\frac{1}{k_F\xi} + \cos 4\theta_{\k_i} \sin \frac{1}{k_F\xi} \right) \,.
    \label{eq:SelfE_nem}
\end{equation}
From \eqref{eq:SelfE_nem}, we note that the glassy nematic order leads to `cold-spots' on the Fermi surface at $\theta_\k = \pm \pi/4, \pm 3\pi/4$ (Fig.~\ref{fig:headimage}d). 
The scattering rate at the cold spots $1/{\tau_{c}} \sim (J^2/\epsilon_F) (k_F\xi)^{-1}$
is suppressed by a factor of $(k_F\xi)^{2}$ relative to the scattering rate at generic points on the Fermi surface, given by $1/\tau_a \sim (J^2/\epsilon_F) (k_F \xi)$ \footnote{Note that the cold spot scattering rate $1/\tau_c$ obeys this scaling relation irrespective of the disorder strength, i.e., whether or not $J \lesssim \frac{v_F}{\xi}$. This is because at the cold-spots fermions couple weakly to the order parameter. The antinodal scattering rate on the other hand depends on the strength of $J$ and saturates to $v_F/\xi$ for strong disorder strengths}.
This suppression arises because of the vanishing coupling to the nematic order at these wave-vectors. 
The scattering rate is not precisely zero because the nodes are smeared by the inverse correlation length $\xi^{-1}$.

\begin{figure}[ht]
    \centering
    \includegraphics[width=.9\columnwidth]{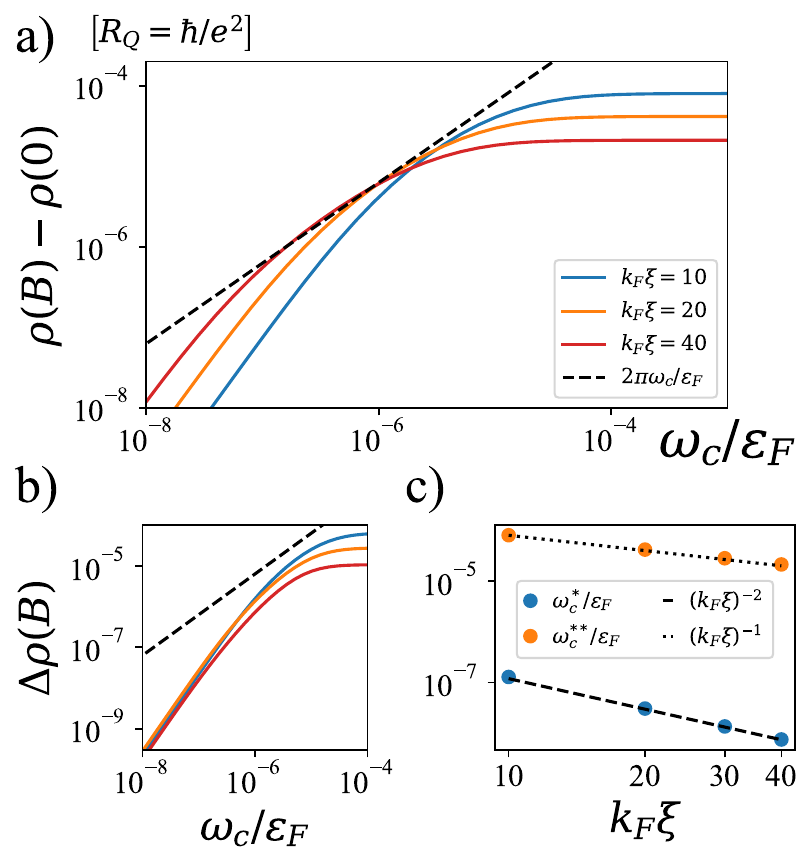}
    \caption{(a) $\Delta \rho(B)$ versus $\omega_c$ for various nematic correlation lengths $\xi$, obtained by numerically solving \eqref{eq:BEmod_2} for $\tau_\mu^{-1} = 10^{-7} \epsilon_F$ and $J = 0.025 \epsilon_F$.
    (b) Absence of LMR when the potential disorder is increased keeping other parameters fixed, such that $\tau_\mu^{-1} = 10^{-4} \epsilon_F$.
    In contrast to model I, there is no scaling collapse upon rescaling by the zero-field resistance $\rho(0)$.
    (c) Numerical evidence for power-law scaling of the lower and upper cross-over fields $\omega_c^{*}$ and $\omega_c^{**}$ with $k_F \xi$.}
    \label{fig:model2}
\end{figure}

It is important to distinguish the above scattering times from the momentum relaxation time that controls transport. 
Because of the large length-scale $\xi$, the disorder effects 
predominantly forward scattering, which is inefficient in relaxing momentum and current. 
Consider an initial non-equilibrium momentum distribution created by an applied electric field. To relax the current, the momentum distribution needs to spread over the entire Fermi surface, while each collision changes it only by a small angular fraction of the Fermi surface $1/(k_F\xi)$.
If we considered each collision as a step in a random walk of a quasiparticle around the Fermi surface, then approximately $N_s \sim (k_F\xi)^2$ such steps would be needed to spread the momentum  distribution around the Fermi surface and thereby relax the current.

In the case of scattering on the glassy nematic order, the cold spots on the Fermi surface present a special bottleneck for momentum relaxation.
To understand the origin of this bottleneck, consider a quasiparticle at a cold spot. 
Of the $N_s$ scattering events it needs to move an O(1) angle away from the cold-spot, approximately $\sqrt{N_s}$ scattering events take place in the vicinity of the cold-spot \footnote{The exact scaling, with the numerical pre-factor, is derived in the SI.}. 
Therefore, the total time spent near the cold-spot is given by $\tau_c \sqrt{N_s} \approx (\epsilon_F/J^2) (k_F \xi)^2$.
In contrast, the time spent away from the cold-spot is upper bounded by $\tau_a N_s \sim (\epsilon_F/J^2) (k_F \xi) \ll \tau_{c} \sqrt{N_s}$.
Therefore, the large enhancement of scattering time at the cold-spots only allows the momentum distribution to spread over the Fermi surface at a timescale $\tau_{re}$,
\begin{equation}
    \tau_{re} \simeq \tau_c \sqrt{N_s} = \frac{\epsilon_F}{J^2} (k_F\xi)^2 \,.
    \label{eq:taure}
\end{equation}
Once a magnetic field is turned on quasiparticles begin rotating around the Fermi surface at angular frequency $\omega_c$, thus providing a new mechanism to escape the cold-spots.
When $\omega_c \gtrsim \omega_c^* = 1/\tau_{re}$, the bottleneck in momentum relaxation is replaced with the angular rotation speed $\omega_c$. 
As a result, the relaxation rate of the non-equilibrium momentum distribution is set by $\omega_c$, and LMR sets in. 

We verify the above predictions through a numerical solution to the DC Boltzmann equation with the collision integral arising from the nematic disorder
\begin{equation}
\begin{split}
    & \frac{e v_F \hat{\k} \cdot \mathbf{E}}{2\pi} + \omega_c \nabla_{\theta_\k} \delta N_{\hat{\k}} = - \frac{1}{\tau_\mu} \delta N_{\hat{\k}} \\
    & + \int_{\hat{\k'}} \frac{\pi J^2}{\epsilon_F} (k_F\xi)^2 \cos^2 2\theta_{\hat{\k}+\hat{\k'}} e^{-\xi^2 k_F^2 |\hat{\k}-\hat{\k'}|^2} (\delta N_{\hat{\k'}} - \delta N_{\hat{\k}} ) \,.
    \label{eq:BEmod_2}
\end{split}
\end{equation}
The magnetoresistance $\Delta \rho(B)$ obtained from solving this equation is shown in Fig.\ref{fig:model2},
and it indeed shows LMR at intermediate magnetic fields, with the universal relaxation rate $\tau_B^{-1} \approx \tilde{\mu}_B B$.
As in model I, there are two  crossovers in the behavior of the magnetoresistance. 
The first crossover, from $B^2$ to $B$ linear, occurs at the aforementioned $\omega_c^* = 1/\tau_{re}$. The magnetoresistance saturates to a constant above the higher crossover scale $\omega_c^{**} = \tau_a^{-1}/(k_F\xi)^2\sim k_F\xi \,\omega_c^*$ (Fig.~\ref{fig:model2}c).

The saturation of the magnetoresistance at high magnetic fields can also be understood semi-classically. 
At high fields, the quasi-particles begin to orbit around the Fermi surface faster than they can relax momentum.
Therefore, transport is dominated by fast momentum relaxation away from the cold-spots.
Given the scattering rate $\tau_a^{-1}$ on a generic point on the Fermi surface, and $N_s \sim (k_F \xi)^{2}$ scattering events required to relax momentum, the fast momentum relaxation rate is simply $(N_s \tau_a)^{-1}$.
Consequently, the magnetoresistance saturates when $\omega_c \gtrsim \omega_c^{**} = (N_s \tau_a)^{-1}$.  
In the weak-coupling limit $J \lesssim v_F/\xi$, the Fermi's golden rule result in \eqref{eq:SelfE_nem} implies $1/\tau_a \lesssim (k_F \xi)^{-3} \epsilon_F$, and thus $\omega_c^{**} \sim (J^2/\epsilon_F) ( k_F \xi)^{-1}\lesssim (k_F\xi)^{-3} \epsilon_F$.
For stronger coupling strengths $J \gtrsim v_F/\xi$, we expect $1/\tau_a \sim v_F/\xi$, and accordingly $\omega_c^{**} \sim (k_F\xi)^{-3} \epsilon_F$.

Finally, we remark that potential disorder does not affect either the lower or upper bound of LMR in Model 2; rather, it determines whether LMR is observed or not.
The mechanism for LMR relies on the existence of a bottleneck of momentum relaxation at the cold-spots.
Therefore, if the relaxation rate from potential disorder $1/\tau_\mu$ is significantly greater than the momentum relaxation rate $1/\tau_{re}$ from the nematic disorder, LMR is lost.
Consequently, LMR due to glassy nematicity requires weak potential-disorder scattering, quantified by $1/\tau_\mu \lesssim (J^2/\epsilon_F)(k_F\xi)^{-2}$: nevertheless, within this range of $1/\tau_\mu$ LMR can still be robust to changes in residual resistivity \cite{FeSe}.
Note that this requirement of weak disorder also implies that there is no Kohler scaling collapse of the magnetoresistance for glassy nematicity as the residual resistivity $\rho(0)$ gets large, in contrast to glassy density waves \cite{kohler1938magnetischen}.

\textit{Connection to Strange Metals.-}
Thus far, our analysis has focused entirely on magnetoresistance at low $T$.
However, in experiments, LMR is typically observed in conjunction with strange metallic behavior where the zero-field resistivity scales linearly with temperature \cite{LSCO,2009Sci...323..603C,Ghiotto,Hayes,Jaoui}.
This motivates us to ask: what is the effect of non-zero temperatures on magnetoresistance in our microscopic models?
We address this question phenomenologically by assuming a Planckian scattering rate $1/\tau_{\mu}\approx k_B T$ \cite{Bruin804} and explore the connection to experiments. 

High $T_c$ superconductors like cuprates and pnictides typically exhibit a $B/T$ scaling of magnetoresistance at finite temperatures \cite{Hayes}, which is equivalent to Kohler's scaling following the replacement $1/\tau_\mu \to k_B T$. This behavior is consistent with glassy density waves (model I) as a possible mechanism for LMR in these materials.
Within this model (and taking $1/\tau_\mu \approx k_B T$, LMR should be observed above the crossover field $\tilde\mu_B B^* \approx k_B T/ (k_F \xi)$.
For moderate disorder strength of $k_F\xi \sim 10$, and $m_f = 4m_e$ \cite{LSCOmass2}, we find that $B^*/T \simeq 0.5$ Tesla per Kelvin, which is in reasonable agreement with experiments on cuprates and pnictides \cite{LSCO,Hayes}.
Further, the observed lack of saturation of magnetoresistance in cuprates up to fields of $60$ Tesla \cite{LSCO,Hayes} is consistent with our prediction of an upper critical field of $\omega_c^{**}/\mu_B \approx \epsilon_F/[\mu_B (k_F \xi)^2] \gtrsim 100$ Tesla for $k_F \xi \sim 10$.
Finally, we note that in generating this $B/T$ scaling as observed in these experiments, we are invoking Planckian scattering without residual resistivity. 
While certain cuprates do have weak residual resistivity \cite{barivsic1,barivsic2}, the presence of significant residual resistivity due to disorder can affect this scaling, which remains an important question for future work.

In strongly correlated moir\'e materials such as twisted bilayer graphene, magnetoresistance does not obey $B/T$ scaling in the strange metallic phase \cite{Jaoui}. 
Further, LMR is suppressed by increasing temperature to a few Kelvins, and is replaced by Fermi-liquid like $B^2$ magnetoresistance \cite{Jaoui}.
These features are consistent with a glassy nodal order parameter (model II), such as strain or inter-valley coherence \cite{BCZ_PRL,Liu,Dan_PRL}, as the mechanism for LMR in these materials. 
For typical values $k_F\xi \sim 10$, $\epsilon_F \approx 30$ meV \cite{BM,pomeranchukTBLG} and $J \sim \epsilon_F$ \cite{Nick_PRX}, the crossover temperature scale, above which LMR is suppressed, is $T^* =  (J^2/\epsilon_F) (k_F\xi)^{-2} \approx 4 K$, consistent with experiments on magic angle graphene \cite{Jaoui}.

\textit{Summary and Outlook.-}
In this paper, we demonstrated with two microscopic models that $B$ linear magnetoresistance is a ubiqutous phenomenon in systems proximate to order, where the order parameter either has a finite wavevector or has nodes on the Fermi surface.
In particular, both models exhibited a $B$ linear relaxation rate with a coefficient given by the effective Bohr magneton as observed in recent experiments \cite{Hayes,LSCO,Ghiotto,Jaoui}.
The first model that we presented for LMR -- with glassy CDW order -- provides a microscopic model for the mechanism suggested in Ref.\cite{Hussey}.
In contrast to previous microscopic realizations of anisotropic hot-spot scattering, our results do not rely on the existence of long-range density-wave order \cite{Feng,Maksimovic,Koshelev2}, or thermal fluctuations of the order parameter \cite{Koshelev1,Rosch} - but rather on static domains pinned by disorder.
In addition, we demonstrated through the second model -- with glassy nodal order -- that an entirely different mechanism, namely, releasing the bottleneck of momentum relaxation at cold-spots, could give rise to LMR.
While LMR in the latter model is sensitive to potential disorder, as characterized by the residual resistivity, LMR survives upto a critical strength of disorder similar to observations in doped FeSe proximate to a nematic critical point \cite{FeSe}.

Although both models that we presented in this paper were comprised of some type of charge order, we expect the same physics to be present even for spin order such as spin density waves or spin nematic order.
Further, while our computations were carried out for single-band model, they can be generalized to materials with multiple Fermi pockets, such as pnictides or moir\'e graphene. 
In such multi-band systems, the total conductivity depends on the conductance of each band, and inter-band scattering processes.
Provided that the interband scattering processes are weak, we expect LMR to be robust in systems with glassy density waves as long as each Fermi pocket has a hotspot.
In this case, the electron is forced to encounter a hotspot in course of its cyclotron motion on the Fermi pocket(s).
By contrast, if a Fermi pocket has a nodal point and other pockets do not, it will dominate the conductivity, as faster momentum relaxation in the pockets without nodes increases their resitivity. 
Therefore, we expect LMR to be robust for glassy nodal order so long as one of the Fermi pockets contain a node.

Our work opens the door to several new directions. 
Our semiclassical approach sets the stage to investigate the origin of LMR in strange metal phases without sharply defined quasiparticles \cite{Chowdhury1,Chowdhury2,Patel2,Patel3} via quantum Boltzmann equations \cite{KimLeeWen}.
A natural next step is to include dynamical order parameter fluctuations that can potentially play an important role in transport in the vicinity of quantum critical points, where one often observes strange metallic behavior.
Additionally, the effect of correlated pairing disorder on magnetoresistance can be studied by an appropriate generalization of our formalism, and is discussed in a companion paper \cite{Kim2024}.
Finally, the ability to tune relaxation processes through magnetic fields may lead to field-tunable thermoelectric effects, and is left for future work.

\acknowledgements{We would like to thank Erez Berg, Zhehao Dai, Kedar Damle, Unmesh Ghorai and Vikram Tripathi for helpful discussions and comments. E.A. acknowledges support from the Simons Investigator award.}

% Bibliography
\bibliography{ref}

\appendix

\begin{widetext}
\section{Comparison with Experiments}
Table.\ref{tab:comparison} presents Fig.1a) of the main text in table format and provide a comparison between the quasiparticle mass $m_f$ found from quantum oscillation experiments, and that found from a Drude analysis.

\renewcommand{\arraystretch}{1.5}
\begin{table}[h]
    \centering
    \begin{tabular}{|c|c|c|c|}
    \hline
    Material & \thead{Carrier Density (m$^{-2}$)} & \thead{$m_f/m_e$ from \\ slope of LMR} & \thead{$m_f/m_e$ from \\ Quantum Oscillations} \\ \hline
    LSCO & $2.0 \times 10^{18}$ \cite{LSCOdensity}  & 5.4 & 4 \cite{LSCOmass2} \\  \hline
    BaFe$_2$(As$_{1-x}$P$_x$)$_2$ & $4.6\times10^{18}$ \cite{Pnictidemass} & 4.5 & 3.2 \cite{Pnictidemass} \\ \hline
    tWSe$_2$ & $5.0 \times 10^{15}$ \cite{Ghiotto} & 0.71 & 0.4 \cite{Ghiotto} \\ \hline
    \thead{MATBG \\ $(\nu = -2.8)$} & $5.0 \times 10^{15}$ & 0.18 & 0.3 \cite{Jaoui} \\ \hline
    \end{tabular}
    \caption{Comparison between the quasiparticle mass found through quantum oscillation experiments and that from the slope of LMR for various materials exhibiting LMR.}
    \label{tab:comparison}
\end{table}

\section{Methods: Derivation of the Boltzmann Equations}
\label{app:BE}

\subsection{Derivation via a quantum Boltzmann approach}
In this appendix, we derive the Boltzmann equations that we use in the main text.
Our starting point is the quantum Boltzmann equation in the DC limit which reads \cite{KimLeeWen},
\begin{equation}
    [A_G(\k,\omega)]^2 \, \Im[\Sigma_R(\k,\omega)] \, \left( \frac{e \k \cdot \mathbf{E}}{m_f} \right) {f_0}'(\omega) + \left(\frac{e \k \times B\hat{z}}{m_f} \right) \cdot \nabla_{\k} G^<(\k,\omega) = \Sigma^> G^< - \Sigma^< G^> \,.
    \label{eq:appBE}
\end{equation}
In the above equations, $G^{>,<}$ denote the forwards and backwards electron Green's functions, and $\Sigma^{>,<}$ are the corresponding self-energies (as defined in Ref.\cite{KimLeeWen}), and we have suppressed the $(\k,\omega)$ indices on the RHS for clarity.
$A_G$ on the other hand stands for the electron spectral function, defined in terms of the imaginary part of the retarded electron Green's function as $A_G(\k,\omega) = -2\Im\{G_R(\k, \omega)\}$.
The left hand side of \eqref{eq:appBE} denotes electromagnetic force felt by the quasi-particle; the right hand side on the other hand, is called the collision integral and is composed of the scattering processes experienced by the quasi-particles.

We use the fact that the electron spectral function $A_G$ is sharply peaked near the Fermi surface at low energy $\omega$, and perform a change of variables from $\k$ to $\hat{k}$ and $\Delta k = |\k|-k_F$.
Let us define $f(\hat{k},\omega)$ as the generalized distribution function for the density of electrons with momentum in the direction $\hat{k}$ and frequency $\omega$. 
It satisfies
\begin{equation}
f(\hat{k},\omega) = v_F \int_{\Delta k} G^< \big((k_F + \Delta k)\hat{k},\omega \big) \,, \ \
1-f(\hat{k},\omega) = v_F \int_{\Delta k} G^> \big((k_F + \Delta k)\hat{k},\omega \big)
\label{eq:appf}
\end{equation}
For convenience of notation, we define $\int_{\Delta k} = \int \frac{d \Delta k}{2\pi}$.
At equilibrium, $f(\hat{k},\omega)$ is simply given by the Fermi-Dirac distribution, $f_0(\hat{k},\omega) = \frac{1}{1+e^{\beta\omega}}$.
We shall denote the deviation from the equilibrium value by $\delta f(\hat{k},\omega)$.

Upon integrating the left hand side of \eqref{eq:appBE}, we get, 
\begin{equation}
\begin{split}
    & \int_{\Delta k} [A_G(\k,\omega)]^2 \, \Im[\Sigma_R(\k,\omega)] \, \left( \frac{e \k \cdot \mathbf{E}}{m_f} \right) {f_0}'(\omega) + \left(\frac{e \k \times B\hat{z}}{m_f} \right) \cdot \nabla_{\k} G^<(\k,\omega) \\
    & \quad \simeq \int_{\Delta k} \frac{4 \Gamma^3} {\big((\omega - v_F \Delta k)^2 + \Gamma^2 \big)^2}
    e v_F\hat{k} \cdot \mathbf{E} {f_0}'(\omega) + \left(\frac{e \hat{k} \times B\hat{z}}{m_f} \right) \cdot \nabla_{\theta_k} G^<(\k,\omega) = e \hat{k} \cdot \mathbf{E} f_0'(\omega) + \frac{\omega_c}{v_F} \nabla_{\theta_k} \delta f(k,\omega)
    \label{eq:RHS}
\end{split}
\end{equation}
Here, $\Gamma$ denotes the imaginary part of the self-energy, $\Im[\Sigma_R(\k,\omega) ]$.

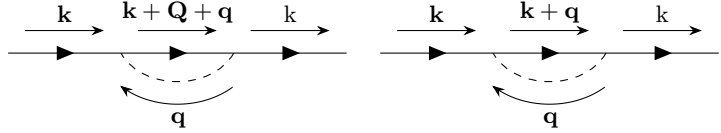
\begin{figure}[h]
    \centering
    \begin{tikzpicture}[baseline=(a)]
    \begin{feynman}[medium]
    \vertex (a);
    \vertex [right=of a] (b);
    \vertex [right=of b] (c);
    \vertex [right=of c] (d);
    \diagram* {
    (a) -- [fermion, momentum = $\k$] (b) -- [fermion, momentum = $\k+\Q+\q$] (c) -- [fermion, momentum = k] (d),
    (c) --[scalar, out = -120, in = -60, looseness = 1, momentum = $\q$] (b)
    };
    \end{feynman}
    \end{tikzpicture}
    \quad 
    \begin{tikzpicture}[baseline=(a)]
    \begin{feynman}[medium]
    \vertex (a);
    \vertex [right=of a] (b);
    \vertex [right=of b] (c);
    \vertex [right=of c] (d);
    \diagram* {
    (a) -- [fermion, momentum = $\k$] (b) -- [fermion, momentum = $\k+\q$] (c) -- [fermion, momentum = k] (d),
    (c) --[scalar, out = -120, in = -60, looseness = 1, momentum = $\q$] (b)
    };
    \end{feynman}
    \end{tikzpicture}
    \caption{Self-energies from the glassy disorder for model I (\textit{left}), and model II (\textit{right}). Solid lines denote the electron propagator, while the dotted lines indicate disorder scattering.}
    \label{fig:appFD}
\end{figure}

Let us now compute the right hand side of \eqref{eq:appBE}.
We first focus on the collision integral from scattering caused by the glassy order parameter.
To this end, we need to evaluate the self-energy from these scattering processes.
The Feynman diagrams of the self-energies from the glassy order parameters are each given in Fig.\ref{fig:appFD}.

Evaluating the Feynman diagrams, we find that the self-energy from the glassy disorders are each given as,
\begin{subequations}
\label{eq:appSigma}
\begin{align}
    \Sigma_{h}^>(\k,\omega) &= \int_{k'} 2\pi J^2 \xi^2 e^{-\xi^2 |\k-\k' \pm \Q|^2} G^>(\k',\omega)
    \label{eq:appSigmah} \,, \\
    \Sigma_{c}^>(\k,\omega) &= \int_{q} 2\pi J^2 \xi^2 e^{-\xi^2 q^2} \big(\cos (k_x + q_x/2) - \cos (k_y + q_y/2) \big)^2 G^>(\k+\q,\omega) \label{eq:appSigmac} \\
    & \simeq \int_{\q}^{q < \frac\pi\xi} 2\pi J^2 \xi^2 \cos^2 2\theta_{\k+\q/2} G^>(\k+\q,\omega)
    \nonumber \,.
\end{align}
\end{subequations}
In ~\eqref{eq:appSigma}, $\Sigma_h$ is the self-energy from model I dynamics, where the order parameter for the glassy disorder possesses a finite wave-vector $\mathbf{Q}$.
On the other hand, $\Sigma_c$ is the self-energy from model II dynamics, where the order parameter has nodes on the Fermi surface.

In turn, the self-energies of ~\eqref{eq:appSigma} lead to the following collision integral: 
\begin{equation}
    \Sigma^> G^< - \Sigma^< G^> = \begin{cases}
    \int_{\k'} 2\pi J^2 \xi^2 e^{-\xi^2 |\k-\k' \pm \mathbf{Q}|^2} \Big( G^>(\k,\omega) G^<(\k',\omega) - G^<(\k,\omega) G^> (\k',\omega) \Big) ~~~ \textrm{ for Model I} \\ \\
    \int_{\k'} 2\pi J^2 \xi^2 \cos^2 2\theta_{\left(\frac{\k+\k'}{2}\right)} \, e^{-\xi^2 |\k-\k'|^2} \Big( G^>(\k,\omega) G^<(\k',\omega) - G^<(\k,\omega) G^> (\k',\omega) \Big) ~~~ \textrm{ for Model II}
    \end{cases}
\end{equation}

Integrating out the momenta perpendicular to the Fermi surface, and expanding to the lowest order in $\delta f$, we obtain,
\begin{subequations}
\label{eq:CI_glassy}
\begin{align}
    & \int_{\hat{k'}} \frac{2\pi J^2 \xi^2 k_F}{v_F^2} e^{-\xi^2 |k_F \hat{k}-k_F \hat{k'} \pm \Q|^2} \Big(\delta f(\hat{k'},\omega) - \delta f(\hat{k},\omega) \Big) \textrm{ for Model I} \\
    & \int_{\hat{k'}} \frac{2\pi J^2 \xi^2 k_F}{v_F^2} \cos^2 2\theta_{\frac{k+k'}{2}} e^{-\xi^2 |k_F \hat{k}-k_F \hat{k'}|^2} \Big(\delta f(\hat{k'},\omega) - \delta f(\hat{k},\omega) \Big) \textrm{ for Model II}
\end{align}
\end{subequations}

Finally, in the case of uncorrelated potential disorder, it leads to the following self-energy:
\begin{equation}
    \Sigma_{\mu}^{>}(\k,\omega) = g^2 \int_{k'} G^>(k',\omega) \simeq m_f g^2 \big(1-f_0(\omega) \big) \,.
\end{equation}
If we define the scattering rate due to potential disorder as $\tau_\mu^{-1} = m_f g^2$, the collision integral from such a self-energy is,
\begin{equation}
    \int_{\Delta k} \Sigma_\mu^>(\k,\omega) G^<(\k,\omega) - \Sigma_\mu^<(\k,\omega) G^>(\k,\omega)  = -\frac{1}{\tau_\mu v_F} \delta f(\hat{k},\omega) \,,
    \label{eq:CIdd}
\end{equation}

Putting Eqs.~\eqref{eq:RHS},\eqref{eq:CI_glassy},\eqref{eq:CIdd} together, we arrive at the following set of quantum Boltzmann equations:
\begin{equation}
\begin{split}
    &e \hat{k} \cdot \mathbf{E} f_0'(\omega) + \frac{\omega_c}{v_F} \nabla_{\theta_k} \delta f(\hat{k},\omega) \\ &=
    \begin{cases}
    \int_{\hat{k'}} \frac{2\pi J^2 \xi^2 m_f}{v_F} e^{-\xi^2 |k_F \hat{k}-k_F \hat{k'} \pm \Q|^2} \Big(\delta f(\hat{k'},\omega) - \delta f(\hat{k},\omega) \Big) - \frac1{\tau_\mu v_F} \delta f(\hat{k},\omega) \textrm{ for Model I} \\
    \\
    \int_{\hat{k'}} \frac{2\pi J^2 \xi^2 m_f}{v_F} \cos^2 2\theta_{\frac{k+k'}{2}} e^{-\xi^2 |k_F \hat{k}-k_F \hat{k'}|^2} \Big(\delta f(\hat{k'},\omega) - \delta f(\hat{k},\omega) \Big) - \frac1{\tau_\mu v_F} \delta f(\hat{k},\omega) \textrm{ for Model II}
    \end{cases}
\end{split}
\end{equation}
Integrating both sides with regards to frequency, we arrive at the semiclassical Boltzmann equations for the quasiparticle density $\delta N(\hat{k}) = \int_\omega \delta f(\hat{k},\omega)$ around the Fermi surface.
\begin{equation*}
\begin{split}
    \frac{e v_F \hat{k} \cdot \mathbf{E}}{2\pi} + \omega_c \nabla_{\theta_k} \delta N(\hat{k}) &= \int_{\hat{k'}} \frac{\pi J^2}{\epsilon_F} (k_F\xi)^2 e^{-\xi^2 |k_F \hat{k}-k_F \hat{k'} \pm \Q|^2} \Big(\delta N(\hat{k'}) - \delta N(\hat{k}) \Big) - \frac1{\tau_\mu} \delta N(\hat{k}) \textrm{ for Model I} \\
    \frac{e v_F \hat{k} \cdot \mathbf{E}}{2\pi} + \omega_c \nabla_{\theta_k} \delta N(\hat{k})  &= \int_{\hat{k'}} \frac{\pi J^2}{\epsilon_F} (k_F\xi)^2 \cos^2 2\theta_{\frac{k+k'}{2}} e^{-\xi^2 |k_F \hat{k}-k_F \hat{k'}|^2} \Big(\delta N(\hat{k'}) - \delta N(\hat{k}) \Big) - \frac1{\tau_\mu} \delta N(\hat{k})\textrm{ for Model II}
\end{split}
\end{equation*}
These are the Boltzmann equations used in the main text.

\subsection{Semi-classical derivation using Fermi's golden rule}
In this subsection, we derive the semiclassical Boltzmann equation through an alternate approach, using Fermi's golden rule.
This also provides a derivation of the scattering rates from Fermi's golden rule, that we used for our heuristic arguments in the main text. 

According to Fermi's golden rule, the rate at which a quasiparticle at momentum $\k_i$ scatters to momentum $\k_f$ by the glassy order is given by,
\begin{subequations}
\begin{align}
    \textrm{Model I}: \Gamma_{\k_i \rightarrow \k_f} &= 2\pi \left| \bra{\k_f} H_{dw} \ket{\k_i} \right|^2 \delta(\epsilon_{\k_f}- \epsilon_{\k_i}) \\
    &= \frac{2\pi^2 J^2 \xi^2}{v_F} e^{-\xi^2 q^2} \delta(|\k_f|-|\k_i|) \,, \textrm{ where } \q = \Delta \k \pm \Q
    \label{eq:appGamma1} \\
    \textrm{Model II}: \Gamma_{\k_i \rightarrow \k_f} &= 2\pi \left| \bra{\k_f} H_{nem} \ket{\k_i} \right|^2 \delta(\epsilon_{\k_f}- \epsilon_{\k_i}) \\
    &= \frac{2\pi^2 J^2 \xi^2}{v_F} \cos^2 2\theta_{\frac{\k_i+\k_f}{2}} e^{-\xi^2 |\k_i-\k_f|^2} \delta(|\k_f|-|\k_i|)
    \label{eq:appGamma2}
\end{align}
\end{subequations}
Integrating \eqref{eq:appGamma1} with regards to the final momentum $k_f$ gives the lifetime of a quasiparticle at momentum $k_i$ on the hotspot in model 1.
It is given as,
\begin{equation}
\begin{split}
    \frac{1}{\tau_h} &= \int_{\k_f} \Gamma_{\k_i \rightarrow \k_f} = \int \frac{k_F d \hat{k}_f}{(2\pi)^2} \frac{2\pi^2 J^2 \xi^2}{v_F} e^{-\xi^2 q^2} \textrm{ where } q = k_F (\hat{k}_i - \hat{k}_f) - \Q \\
    &\simeq \frac{J^2 \xi^2 k_F}{v_F} (k_F\xi)^{-1}
\end{split}
\end{equation}
Here, in the second step we have approximated the gaussian function as a $\Theta(1/\xi-|q|)$.

Similarly, integrating \eqref{eq:appGamma1} with regards to the final momentum $\k_f$ gives the lifetime of a quasiparticle at momentum $\k_i$ for model 2.
\begin{equation}
\begin{split}
    \frac{1}{\tau_{k_i}} &= \int_{\k_f} \Gamma_{\k_i \rightarrow \k_f} = \int \frac{k_F d \hat{k}_f}{(2\pi)^2} \frac{2\pi^2 J^2 \xi^2}{v_F} \cos^2 2\theta_{\frac{\k_i+\k_f}{2}} e^{-\xi^2 |\k_i - \k_f|^2} \\
    &\simeq \int_{\theta_{\k_i}-(k_F\xi)^{-1}}^{\theta_{\k_i}+(k_F\xi)^{-1}} k_F d\theta_{\k_f} \frac{J^2 \xi^2}{2 v_F} \cos^2 2\theta_{\frac{\k_i+\k_f}{2}} = \frac{J^2}{2\epsilon_F} (k_F\xi)^2 \left(\frac{1}{k_F\xi} + \cos 4\theta_{\k_i} \sin \frac{1}{k_F\xi} \right)
\end{split}
\end{equation}

In addition, these scattering rates result in the following semiclassical Boltzmann equation:
\begin{equation}
\begin{split}
    ev_F \hat{k} \cdot \mathbf{E} f_0'(\epsilon_k) + \omega_c \nabla_{\theta_\k} f(\k) &= -\int_{\k'} \Gamma_{\k \rightarrow \k'} \Big( f(\k) \big(1- f(\k') \big) - f(\k') \big(1-f(\k) \big) \Big) \\
    &= \int_{\k'} \Gamma_{\k \rightarrow \k'} \Big( f(\k') - f(\k) \Big)
\end{split}
\end{equation}
Upon integrating out the momentum perpendicular to the Fermi surface,
we ultimately arrive again to the Boltzmann equations of the main text.
\begin{equation*}
\begin{split}
    \frac{e \hat{k} \cdot \mathbf{E}}{2\pi} + \frac{\omega_c}{v_F} \nabla_{\theta_\k} N(\hat{k}) &= -\int_{\hat{k'}} k_F \int_{\Delta k'} \int_{\Delta k} \Gamma_{\k \rightarrow \k'} \Big( f(\k') - f(\k) \Big) \\
    &= \begin{cases}
    & \int_{\hat{k'}} \frac{\pi J^2 \xi^2 m_f}{v_F} e^{-\xi^2 |k_F \hat{k}-k_F \hat{k'} \pm \Q|^2} \Big( N(\hat{k'}) - N(\hat{k}) \Big) \textrm{ Model I} \\
    & \int_{\hat{k'}} \frac{\pi J^2 \xi^2 m_f}{v_F} e^{-\xi^2 |k_F \hat{k}-k_F \hat{k'}|^2} \Big( N(\hat{k'}) - N(\hat{k}) \Big) \textrm{ Model I} \textrm{ Model II}
    \end{cases}
\end{split}
\end{equation*}
Note that we have used the fact that $v_F \int_{\Delta k} f(\k) = N(\hat{k})$.

\section{Robustness of Our Results to the Precise Form of Correlations}
\label{app:exp}
In this section, we demonstrate that our results on LMR are not sensitive to the precise form of the order-parameter correlations, as long as they decay over a length-scale $\xi$ such that $k_F \xi \gg 1$.
To this end, instead of Gaussian correlated order parameters, we consider an order parameter correlation defined by:
\begin{equation}
    \overline{n_\x} = 0 \,, \overline{n_\x n_{\x'}} = \frac{1}{\left(\frac{|\x-\x'|^2}{\xi^2}+1\right)^{3/2}} \Leftrightarrow \overline{n_\q} = 0 \,, \overline{n_\q n_{\q'}} = 2\pi \xi^2 \, \delta_{\q+\q'} e^{-\xi|\q|}
    \label{eq:appdis_exp}
\end{equation}
The above order parameter correlations result in the following modification of the Boltzmann equations for the two models:
\begin{equation}
\begin{split}
    \frac{e v_F \hat{k} \cdot \mathbf{E}}{2\pi} + \omega_c \nabla_{\theta_k} \delta N(\hat{k}) &= \int_{\hat{k'}} \frac{2\pi J^2}{\epsilon_F} (k_F\xi)^2 e^{-\xi |k-k' \pm Q|}  \Big(\delta N(\hat{k'}) - \delta N(\hat{k}) \Big) - \frac1{\tau_\mu} \delta N(\hat{k}) \textrm{ for Model I} \\
    \frac{e v_F \hat{k} \cdot \mathbf{E}}{2\pi} + \omega_c \nabla_{\theta_k} \delta N(\hat{k}) &= \int_{\hat{k'}} \frac{2\pi J^2}{\epsilon_F} (k_F\xi)^2 e^{-\xi|k-k'|} \Big(\delta N(\hat{k'}) - \delta N(\hat{k}) \Big) - \frac1{\tau_\mu} \delta N(\hat{k})\textrm{ for Model II}
    \label{eq:appBE_exp}
\end{split}
\end{equation}
The results on solving \eqref{eq:appBE_exp} is displayed in Fig.\ref{fig:appBE_exp}.
We see qualitatively the same behavior as that seen in the main text where the order parameter correlation is given by a Gaussian.

\begin{figure}[h]
    \centering
    \includegraphics[width = 0.35 \columnwidth]{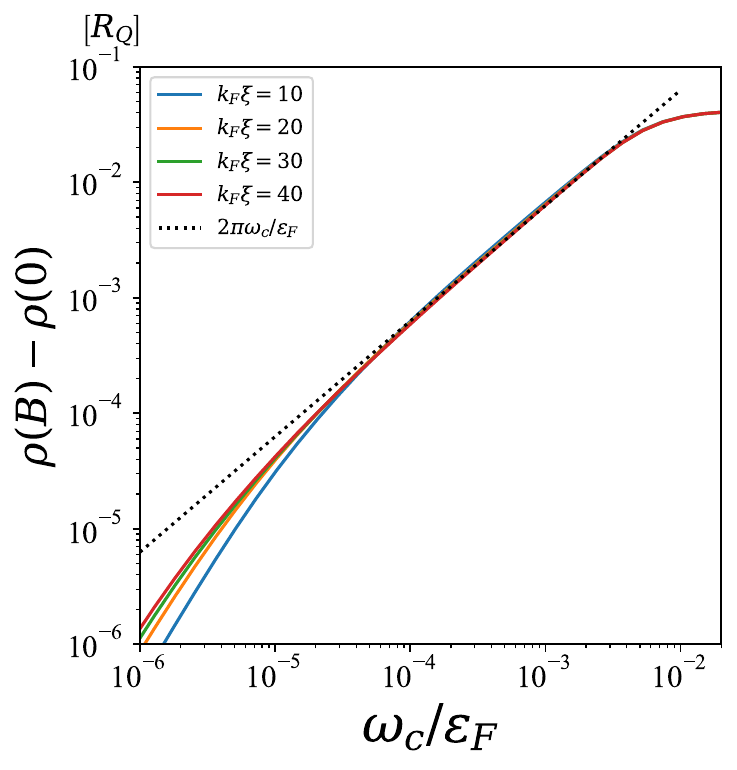}
    \includegraphics[width = 0.35 \columnwidth]{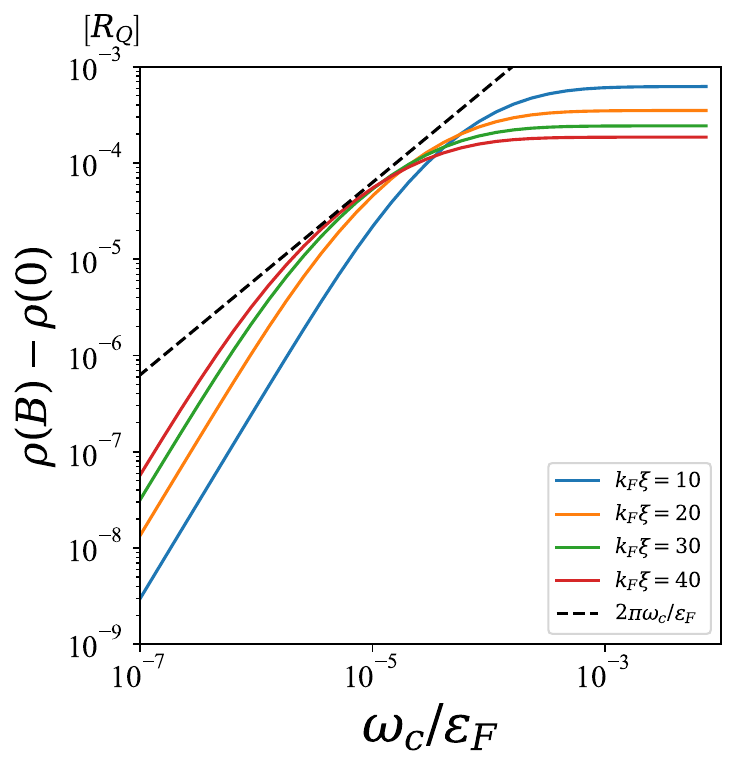}
    \caption{Magnetoresistance for model 1 and 2 obtained for a different form of disorder correlation \eqref{eq:appdis_exp} from solving \eqref{eq:appBE_exp}.}
    \label{fig:appBE_exp}
\end{figure}

\section{Model II Magnetoresistance without Log Scaling the Axes}
In Fig.\ref{fig:RRRmodel2} we plot the magnetoresistance in model 2 without log scaling the axes to highlight the existence of linear magnetoresistance.

\begin{figure}[h]
    \centering
    \includegraphics[width = 0.4\columnwidth]{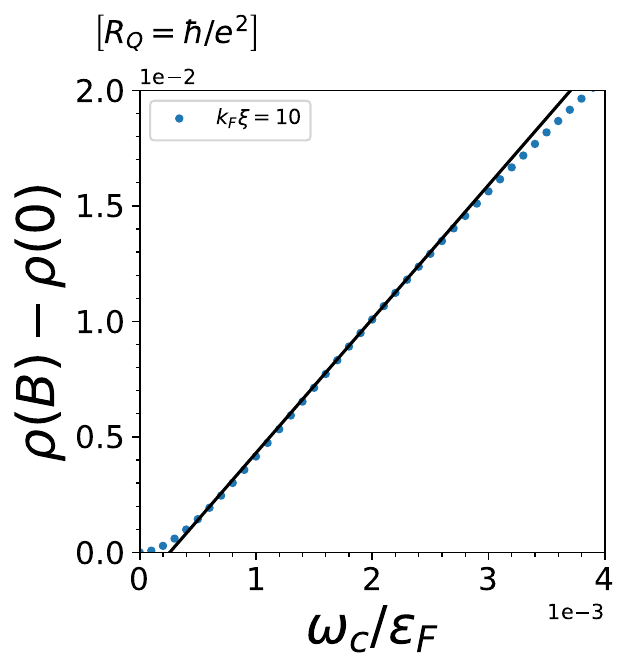}
        \caption{Magnetoresistance for model 2, for $k_F\xi = 10$ and $J \sim \epsilon_F$, plotted without log scaling the axes. Linear magnetoresistance is evident over a decade of magnetic field values, from $\omega_c/\varepsilon_F \approx 2 \times 10^{-4}$ to $\omega_c/\varepsilon_F \approx 3 \times 10^{-3}$.
    }
    \label{fig:RRRmodel2}
\end{figure}

\section{Momentum Relaxation Time in Model II}
\label{app:model2}
In this section, we elaborate on the random walk argument for momentum relaxation time in model II at zero field.  
We model the momentum relaxation process as a random walk of quasiparticles on the Fermi surface.
Each random step occurs when a quasiparticle scatters on the glassy nodal order parameter.
Accordingly, if the quasiparticle is sitting at momentum $\k_i$, the time it takes for the next step is $\tau_{\k_i}$.
After a step (collision) the quasiparticle momentum changes by $1/\xi$, implying that its angular position on the Fermi surface changes by $1/k_F \xi$.
Accordingly, we require an order $N_s = (k_F\xi)^2$ steps for momentum relaxation, for only then the standard deviation in the momentum grows to order $k_F$.

Since $\tau_{\k_i}$ is greatest at the nodes, let us compute the time that a quasi-particle spends at the nodes during its relaxation.
To this end, let us consider a quasiparticle that is placed on the node (which we consider as the origin), and compute the average number of times it passes through the node in $N_s$ steps, out of the total $2^{N_s}$ random sequences that the quasiparticle could take.
We shall denote with $F_k$ the number of sequences in which the quasiparticle reached the origin in $2 k$.
Since doing so requires exactly $k$ forward steps and $k$ backward steps in any order, we have $F_k = \frac{(2k)!}{k! k!} 2^{N_s - 2k}$, where the last factor $2^{N_s - 2k}$ comes from allowing all possible paths after reaching the origin at $2k$ steps.  
Then, summing $F_k$ over $k = 1 \cdots \lfloor{N_s/2 \rfloor}$ counts the total number of times the quasiparticle passes through the origin over all the $2^{N_s}$ random sequences.
For even $N_s$, we have,
\begin{equation*}
    F_{tot} = \sum_{k = 1}^{\frac{N_s}{2}} F_k = \frac{N_s/2 + 1}{2} \frac{(N_s + 2)!}{(N_s/2+1)!(N_s/2+1)!} - 2^{N_s}
\end{equation*}
Consequently, the expected number of times a quasiparticle starting from the origin and undergoing a random walk passes through the origin in $N_s$ steps is given by 
\begin{equation}
\frac{F_{tot}}{2^{N_s}} \simeq \sqrt{\frac{2N_s}{\pi}} -1 + \BigO\left( \frac{1}{\sqrt{N_s}} \right) \text{ for } N_s \gg 1
\end{equation}

\section{Comparison with the Relaxation Time Approximation}
\label{app:WrongApprox}
In the literature \cite{Hussey,2021Natur.595..667G,2022NatPh..18.1420A}, the relaxation time approximation is often used to calculate magnetoresistance via a semi-classical approach.
Notably, in all these models, there are a few hotspots on the Fermi surface with fast momentum relaxation, and the relaxation rate is uniform elsewhere on the Fermi surface. 
In these cases, magnetotransport is similar to our considerations of glassy density wave order, where the fast scattering rate at the hotspot does not determine magnetoresistance. 
Rather, magnetoresistance is either determined by one of the two slower processes - the rate of rotation into a hotspot set by $\omega_c$ (B-linear regime) or the rate of disorder scattering $1/\tau_\mu$ ($B^2$ regime as in conventional Fermi liquid theory).

In this section, we illustrate that the relaxation time approximation needs to be applied carefully when there is a broad anisotropic distribution of scattering rates on the Fermi surface.
As we will show both analytically and numerically, the relaxation time approximation in model II results in the erroneous conclusion that the magnetoresistance scales as $B^{1/3}$.
This incorrect result stems from the fact that quasi-particle relaxation rate can be grossly different from the momentum relaxation rate when the main mode of scattering is forward scattering;
and it is the momentum relaxation rate that sets the transport lifetime.

Let us begin by analytically estimating the conductivity of model II in the main text using the relaxation time approximation.
Recall that the quasi-particle decay rate for a quasiparticle with initial momentum $\k_i$ was given as
\begin{equation}
    \frac{1}{\tau_{\k_i}} \simeq \frac{J^2}{\epsilon_F} (k_F \xi)^2 \left(\frac{\pi}{k_F\xi} + \cos 4\theta_{\k_i} \sin \frac{\pi}{k_F\xi} \right)
    \label{eq:appSigma_2}
\end{equation}
Let us focus on a single node at $\theta_{\k_i} = \frac{\pi}{4}$ and denote with $\phi = \theta_{\k_i} - \frac{\pi}{4}$ the deviation from it.
The decay rate near the nodes is given by,
\begin{equation*}
    \frac{1}{\tau_\phi} \simeq \frac{J^2}{\epsilon_F} (k_F \xi)^2 \left(\frac{\pi^3}{3(k_F\xi)^3} + \frac{8\pi}{k_F\xi} \phi^2 \right)
\end{equation*}

Let us now imploy the Shockley-Chambers tube integral formalism (SCTIF) to evaluate the conductivity \cite{Shockley,Chambers}. It is given by,
\begin{equation}
\begin{split}
    \sigma_{ij} &= \frac1{4\pi^3} \int_{k} \int_{t=0}^{\infty} dt \frac{v_i(0)}{v_F} v_j(-t) \exp \left\{-\int_0^t \frac{dt'}{\tau_0(-t')} \right\}
    \label{eq:SCTIF}    
\end{split}
\end{equation}
Focusing on a single node around $\theta_k = \frac\pi4+\phi$, we have,
\begin{equation}
\begin{split}
    \sigma_{ij} &= \frac{\epsilon_F}{4\pi^3} \int_{\phi} \int_{t=0}^{\infty} \cos \phi \cos (\phi+\omega_ct) \exp\left\{-\int_0^t \frac{dt}{\tau_{\phi+\omega_{c}t}} \right\}
    \label{eq:SCTIF2}
\end{split}    
\end{equation}

We can evaluate this complicated integral by estimating at which time $t_{\phi}$ does the integral inside the exponent of ~\eqref{eq:SCTIF2} is going to reach an $\BigO(1)$ value.
The integral inside the exponent evaluates to
\begin{equation}
    \int_0^t \frac{dt}{\tau_{\phi+\omega_{c}t}} = \frac{J^2}{\epsilon_F} (k_F\xi)^2 \left(\frac{\pi^3}{3(k_F\xi)^3} t + \frac{8\pi}{3 (k_F\xi) \omega_{c}} \Big( (\phi+\omega_{c}t)^3 - \phi^3 \Big) \right)
    \label{eq:SCTIF_integrand}
\end{equation}
For magnetic fields such that $\omega_c \gg \frac{J^2}{(k_F\xi) \epsilon_F}$, we find that \eqref{eq:SCTIF_integrand} becomes of $\BigO(1)$ at a time $t_{\phi}$, given by
\begin{equation}
    t_{\phi} \sim \frac{1}{\omega_c} \left\{ \left(\phi^3 + \frac{\epsilon_F \omega_c}{(k_F\xi) J^2} \right)^{1/3} - \phi \right\}
\end{equation}
In turn, ~\eqref{eq:SCTIF2} can be approximated as,
\begin{equation}
\begin{split}
    \sigma_{ij} & \sim \epsilon_F \int_\phi t_\phi \sim \frac{\epsilon_F}{\omega_c} \int_{\phi} \left(\phi^3 + \frac{\epsilon_F \omega_c}{(k_F\xi) J^2} \right)^{1/3} - \phi \\
    & \sim \frac{\epsilon_F^{5/3}}{J^{4/3} (k_F\xi)^{2/3} \omega_c^{1/3}}
\end{split}
\end{equation}
Where in the second line we have rescaled the integration variable $\phi \rightarrow \left(\frac{\epsilon_F \omega_c}{(k_F\xi) J^2}\right)^{\frac13} \tilde\phi$.
Consequently, this analytical argument demonstrates that the resistance will scale as $B^{1/3}$ within the relaxation time approximation.

\begin{figure}[h]
    \centering
    \includegraphics[width = 0.5 \columnwidth]{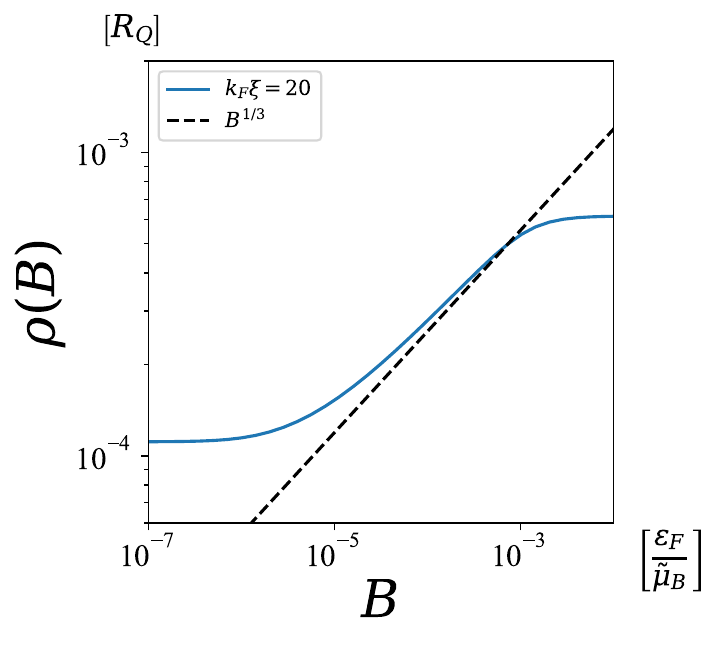}
    \caption{Resistivity versus Magnetic Field plotted in log-log scale, obtained by solving the Boltzmann equations with the relaxation time approximation, \eqref{eq:BE_wrong}. $J = 0.05\epsilon_F$, $k_F\xi = 20$. We observe a resistance that scales as $B^{1/3}$ (Dashed line).}
    \label{fig:BEwrong}
\end{figure}

We also confirm numerically the appearance of this erroneous resistivity that scales as $\rho \propto B^{1/3}$.
In the DC limit, the Boltzmann equation within the relaxation time approximation reads,
\begin{equation}
    \frac{e v_F \hat{k} \cdot \mathbf{E}}{2\pi} + \omega_c \nabla_{\theta_k} \delta N_{\hat{k}} = - \frac{\delta N_{\hat{k}}}{\tau_k} \,.
    \label{eq:BE_wrong}
\end{equation}
The numerical results for the magnetoresistance obtained by solving \eqref{eq:BE_wrong} are illustrated in Fig.\ref{fig:BEwrong}.
We observe a resistivity that scales as $B^{1/3}$ as we analytically derived using the relaxation time approximation.
Note that this conclusion is erroneous, in reality the magnetoresistance scales linearly in $B$ as we showed in the main text.

\end{widetext}

\end{document}